# Gigantic work function in layered AgF$_2$

W. Wegner[1,2], K. Tokár[3,4], J. Lorenzana[5]*, M. Derzsi[2,3]* and W. Grochala[2]

1 College of Inter-Faculty Individual Studies in Mathematics and Natural Sciences, University of Warsaw, Poland
2 Centre of New Technologies, University of Warsaw, Zwirki i Wigury 93, 02089 Warsaw Poland
3 Advanced Technologies Research Institute, Faculty of Materials Science and Technology in Trnava, Slovak University of Technology in Bratislava, J. Bottu 25, 917 24 Trnava, Slovakia
4 Institute of Physics, Slovak Academy of Sciences, 845 11 Bratislava, Slovakia
5 Institute for Complex Systems (ISC), Consiglio Nazionale delle Ricerche, Dipartimento di Fisica, Universita di Roma "La Sapienza" 00185 Rome, Italy

E-mail: jose.lorenzana@cnr.it , mariana.derzsi@gmail.com , w.grochala@cent.uw.edu.pl

**Abstract**

AgF$_2$ is a layered material and a correlated charge transfer insulator with an electronic structure very similar to the parent compounds of cuprate high-Tc superconductors. It is also interesting for being a powerful oxidizer. Here we present a first principles computation of its electronic properties in a slab geometry including its work function for the (010) surface (7.76 eV) which appears to be one of the highest among known materials surpassing even that of fluorinated diamond (7.24 eV). We demonstrate that AgF$_2$ will show a "broken-gap" type alignment becoming electron doped and promoting injection of holes in many wide band gap insulators if chemical reaction can be avoided. Novel junction devices involving p type and n type superconductors are proposed. The issue of chemical reaction is discussed in connection with the possibility to create flat AgF$_2$ monolayers achieving high-Tc superconductivity. As a first step in this direction, we study the stability and properties of an isolated AgF$_2$ monolayer.

This work is dedicated to prof. Michal K. Cyrański at his birthday.

## 1. Introduction

AgF$_2$ is known mostly due to its extremely strong oxidizing and fluorinating properties[1,2]. It belongs to a rather narrow group of about a hundred of Ag(II) compounds[3], which have been extensively studied for the last two decades as possible precursors of high-T$_C$ superconductors[4,5]. The most sought-after feature of AgF$_2$ in this respect are ideally flat sheets, as the expected superexchange in such system exceed values known for cuprates. AgF$_2$ at normal conditions crystallizes in a layered structure in the centrosymmetric *P*bca space group and the AgF$_2$ sheets are structurally and electronically analogous to the CuO$_2$ ones in La$_2$CuO$_4$; however, sheet puckering is much stronger for the former[6]. AgF$_2$ is a narrow band-gap charge-transfer insulator with the calculated fundamental direct gap of *ca.* 1.7 eV [7] – 2.4 eV[8,9]; the experimental estimate is not yet available. The spin-½ transition metal centers reveal strong antiferromagnetic (AFM) interactions with magnetic superexchange constant of ca. 70% of that typical for layered cuprates[8]. As layers in AgF$_2$ are buckled, it was postulated that eliminating or at least reducing this buckling can lead to magnetic exchange constant similar to that of cuprates, or even exceed that value[8,10]. However, it is currently unknown how such flat-sheet AgF$_2$ structures could be prepared.

Since chemical modifications and use of external pressure have failed to deliver desired structural features[11–13], one may contemplate a possibility of fabricating of a monolayer of AgF$_2$ placed on an appropriate substrate material, and then doping it in the field-effect transistor setup. Such approach has proved highly successful in the past in generating superconductivity in oxocuprates[14–16], FeSe[17], SnSe$_2$ [18], tin[19] and other materials[16]. However, before such demanding experiments are performed, one needs to overcome major problems associated with highly aggressive chemical nature of $AgF_2$ [1,2]. It is clear that not all types of substrate materials will be appropriate since many will be susceptible towards electron transfer when in contact with $AgF_2$ monolayer. For example, the recently proposed $SrTiO_3$ (STO)[20,21] is a rather poor choice for a substrate since $AgF_2$ is known to oxidatively destroy all important oxide materials with great ease[22,23]. Use of elemental metals also seems to be out of question, since $AgF_2$ corrodes even platinum electrodes[24]. On the other hand, this strong oxidizing property can be turned into an asset to inject holes in wide-band-gap insulators, a desired property in optoelectronic applications[25].

As a first step to determine the band alignment of $AgF_2$ with respect to other materials, the presently unknown energies of the top of the valence band and bottom of the conduction band in respect to vacuum must be determined. In other words, there is a pressing need to evaluate the work function of $AgF_2$ and its band gap. This will determine which materials will resist electron-transfer at the interface with $AgF_2$ and eventually which materials can benefit from hole injection and provide electron doping. Yet another important question is whether a sheet of $AgF_2$ will spontaneously flatten out while in a monolayer form - or rather it will remain buckled as typical of the bulk structure. The current study aims at providing response to these key questions by examining the structure and electronic properties of $(AgF_2)_N$ (where N=1-11) monolayers and evaluating the work function of $AgF_2$ from density functional theory (DFT) calculations.

## 2. Computational details

The spin polarized DFT calculations [26,27] employing PBEsol exchange-correlation functional [28] were carried out for $AgF_2$ and for several other reference systems, and the value of work function was determined for given slabs, serving as models of surfaces, as outlined in detail in ESI.

## 3. Results and discussion

### 3.1 Geometry analysis

Figure 1 compares selected local geometric parameters of $AgF_2$ layer in: (i) fully optimized bulk $AgF_2$, (ii) a fully optimized single slab (N=1), and (iii) an optimized flat layer slab (N=1) constrained to tetragonal symmetry. The most pronounced structural feature of the fully optimized monolayer in respect to the bulk is the increased puckering. The enhanced puckering is manifested by reduced value of the Ag-F-Ag angle (111.1° in the monolayer vs. 128.9° in the bulk) and it is caused by formation of secondary (axial) intra-layer Ag…F contacts, which substitute the secondary inter-layer Ag…F contacts present in the bulk material. Absence of the inter-layer interactions in the single-layer slab enforces Ag(II) to bind more strongly to intra-sheet F atoms[29]. Puckering of the layers is thus an inherent feature of the $AgF_2$ sheet that dependents on auxiliary interactions (Figure 1). These interactions are completely absent in the hypothetical flat layer and as a consequence it shows the shortest Ag-F distance in this set (2.01 Å)[29]. Its value in the bulk and the fully optimized monolayer is 2.07 Å and 2.08 Å, respectively. Already at N=3 the geometry of the slab resembles greatly that for the bulk system (not shown).

The computed energy penalty to dissociate bulk $AgF_2$ into separate (puckered) monolayers is 0.16 eV per formula unit (FU), and to flatten a single $AgF_2$ layer starting from optimized puckered version requires additional 0.22 eV per FU.

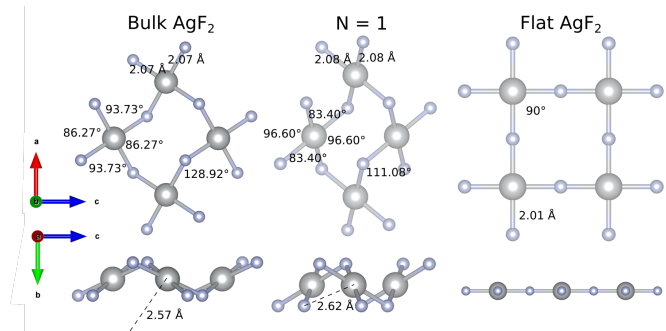

*Figure 1. Comparison of optimized single layer of $AgF_2$: (left) in bulk $AgF_2$; (middle) fully optimized in vacuum, and (right) optimized in vacuum a constrained tetragonal flat sheet model. The shortest secondary contacts are shown with broken lines for bulk $AgF_2$ and optimized monolayer.*

### 3.2 Magnetic properties

Previous study[8] for bulk $AgF_2$ system showed that $AgF_2$ exhibits strong antiferromagnetic interactions between metallic centers. The experimental magnetic superexchange constant, J, for bulk is −70 meV. DFT+U values range between −41 meV for DFT+U method[7], via −56 meV for hybrid HSE06 functional and −71 meV for new generation meta-GGA SCAN functional[8]. Our meta-GGA SCAN calculations yield −71 meV for the bulk system consistently with Ref. [8]. On the other hand, J for the fully optimized puckered layer is computed here to be −7.5 meV, thus considerably reduced as compared to the value calculated for the bulk. This comes from larger bending of the Ag-F-Ag angle within the magnetic interaction pathway, which approaches the value of 90°; this implies frustration of antiferromagnetic superexchange and promoting direct (ferromagnetic) exchange.[30] As expected, the J calculated for a flat $AgF_2$ monolayer is much larger and it reaches −200 meV, in agreement with [8]. This results from opening of the Ag-F-Ag angle to the optimum 180° and simultaneous shortening of the Ag-F bonds as compared to the bulk.

### 3.3 Electronic structure

Electronic density of states (DOS) of bulk AgF$_2$, the fully optimized N=1 layer slab, and the one constrained to tetragonal flat layer are shown jointly in Figure 2 in the energy regions corresponding to the lower Hubbard band (LHB), the valence band and the upper Hubbard band (UHB). The key differences in the shape of the DOS between these systems are: i) the LHB width increases from ca. 0.3 eV for optimized N=1 slab, via 0.4 eV for bulk AgF$_2$, to about 0.7 eV for the flat layer slab; the same sequence is seen in the width of the UHB, which changes from 0.3 eV, via 0.6 eV, up to 1.6 eV, respectively; ii) the band gap at the Fermi level decreases from 1.72 eV, via 1.40 eV, down to 1.27 eV, in the same series.

Consequently, the absolute DOS values just below the Fermi level and at the bottom of the conduction band, as well as steepness of DOS with respect to energy, are much larger for the N=1 layer than for the bulk AgF$_2$, and those in turn is larger than the corresponding values for flat layer system. This implies that charge carrier delocalization (either holes or electrons) will be the easiest for the flat layer while polaronic tendencies will be most pronounced for the optimized single-layer system. Thus, having in mind the possibility of doping charge carriers to AgF$_2$ to induce superconductivity, one should certainly attempt to generate a flat layer system, similar to the flat CuO$_2$ sheets present in most oxocuprate superconductors.

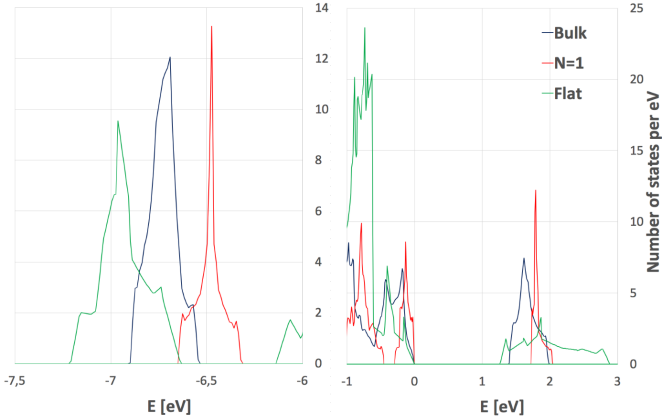

*Figure 2. The calculated total electronic DOS for fully optimized models of bulk AgF$_2$ (Bulk), single layer (N=1), and monolayer constrained tetragonal symmetry (Flat), as shown in the energy regions corresponding to (left) lower Hubbard band, and (right) top of the valence band and conduction (i.e. upper Hubbard) band. Energy of the Fermi level was normalized to zero for all systems.*

## 3.4 Work function

The work function, W, is defined as the energy required to take an electron from the top of the valence band to the vacuum level through a given surface. Sometimes the affinity energy is given which can be computed from our values subtracting the gap energy from the work function[31]. In principle, the work function is sensitive to the details of the charge distribution and relaxations that may occur on the surface of the material. In practice, by fully relaxing the slabs until N=6 we found that the contribution to W of the structural relaxation (labeled opt) is less than 0.1 eV except for the single layer case reported below. Therefore, in the following, unless otherwise specified, we present unoptimized (unopt) results for several number of layers.

Table 1 summarizes the most important results for both unoptimized and optimized N-layer AgF$_2$ slabs. The calculated work function converges with the increasing N to the value of 7.76 eV for unoptimized system (Figure 3). We take this as our best estimate for the (010) surface of bulk AgF$_2$. The work function for an optimized single layer and a flat single layer are, correspondingly, some half eV larger (8.23 eV) and half eV smaller (7.21 eV). Such large change in work function is due to the strong conformational changes with respect to the bulk that are unique of the N=1 systems.

*Table 1. Results obtained from DFT+U calculation for optimized (N opt) and for not optimized (N unopt) N-surface systems of AgF$_2$ constructed from those found in optimized bulk material. We report the band gap (BG), unit cell dimensions (a, c), and work function computed as the difference between vacuum energy and Fermi energy (W=E$_{vacuum}$–E$_F$).*

| N | BG [eV] | a [Å] | c [Å] | W [eV] |
|---|---|---|---|---|
| 1 opt* | 1.723 | 5.728 | 3.726 | 8.232 |
| 1 flat-layer opt | 1.272 | 4.020 | 4.020 | 7.210 |
| 1 unopt | 1.656 | 5.499 | 5.056 | 7.853 |
| 3 unopt | 1.591 | 5.499 | 5.056 | 7.856 |
| 5 unopt | 1.57 | 5.499 | 5.056 | 7.850 |
| 7 unopt | 1.463 | 5.499 | 5.056 | 7.770 |
| 9 unopt | 1.457 | 5.499 | 5.056 | 7.763 |
| 11 unopt | 1.450 | 5.499 | 5.056 | 7.761 |
| Bulk | 1.401 | 5.499 | 5.056 | – |

* crystallographic γ angle is 74.5º (90º for all remaining structures)

The calculated fundamental band gap decreases with the increase of the number of layers in a slab from the value of ca. 1.75 eV computed for N=1 (unoptimized) system to that of ca. 1.40 eV for the N=11 one. The latter value is identical to the one computed here for the bulk AgF$_2$ using the same methodology as for the slabs (1.40 eV). This result also confirms that the N=11 slab system is a fair representation of the bulk AgF$_2$ exposing its surface to vacuum.

The values of the work function found are particularly large. We now analyze the consequences of such high values in relation to possible interfaces between AgF$_2$ layer and other chemical systems. First, to place the values of W reported above for different forms of AgF$_2$ in context, one may consider the series: Ag metal, Ag$_2$F metal[32], AgF semiconductor[33], AgF$_2$ insulator (Figure 4). Work function of silver metal, for its (110), (100), and (111) surfaces, was experimentally estimated from photoelectric spectra to be 4.52

± 0.02 eV, 4.64±0.02 eV, and 4.74 ± 0.02 eV, respectively[34,35]. For comparison, work function for Ag$_2$F (a layered compound with a natural cleavage exposing the (001) surface) was evaluated at 5.54 eV and 5.49 eV, from the threshold value of photoelectric emission and from the photoelectron spectroscopy, respectively[32]. Our own theoretical estimates (ESI) fall close to the experimental value yielding W of 5.24 eV, i.e. only a quarter eV off the experimental value. The value calculated here for the Ag(I) system, AgF, is 5.22 eV, not far from that for that computed for Ag$_2$F. On the other hand, the value of W calculated here for AgF$_2$ is as large as 7.76 eV; this implies a qualitative difference in comparison to the above-mentioned silver systems. Clearly, the partial depopulation of the 4d states upon the Ag(I)→(Ag(II) ionization, and the resulting markedly covalent bonding of Ag(II) cations to fluoride anions[36], lead to the substantial increase of the electron binding energy at fluoride sites, which predominantly contribute to the states at the top of the valence band[8].

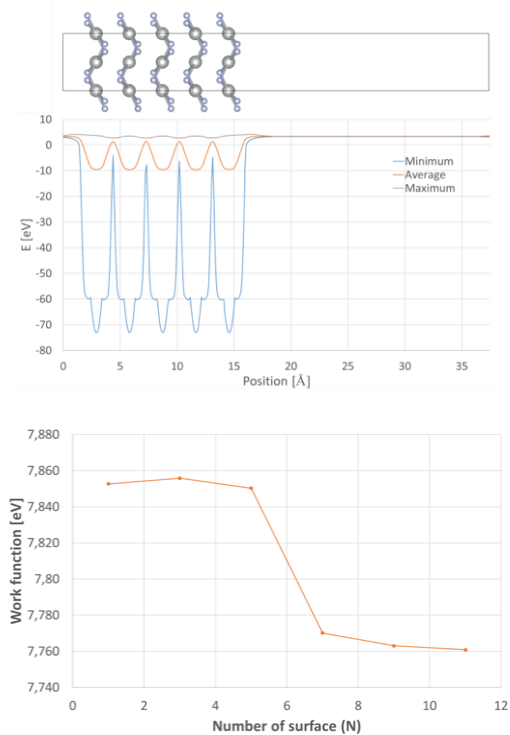

*Figure 3. An exemplary 5-layer (N=5) AgF$_2$ system with the vacuum slab (top left) and the associated electrostatic potential (bottom left). The work function, W, for AgF$_2$ N=1…11 systems with a vacuum slab as calculated from first principle DFT+U calculations with AgF$_2$ sheets preserving geometry found in the bulk crystal (right).*

An even better understanding of the W value for AgF$_2$ is provided by its comparison (Figure 4) to those measured for a broader series of metals, semiconductors and insulators[37–39]. The work function for AgF$_2$ surpasses those measured for two classical oxide electrode materials, i.e. quasi-stoichiometric TiO$_2$ and CuO, by ca. 2.25–2.45 eV. This suggests that if an AgF$_2$ sheet was placed on the surface of those materials, electron transfer between the two would be inevitable. In other words, the surface of TiO$_2$ would be oxidized by introducing holes to O$^{2-}$ states; this, indeed, is observed in reactions between bulk AgF$_2$ and most metal oxides upon heating[22]. According to the affinity model this means that AgF$_2$ will show a so called "broken-gap" alignment with most semiconductors and insulators[31]. How much charge is transferred requires detailed computations of the specific interface and goes beyond of our present scope. In the case of a MgO/AgF$_2$ interface[40] it was shown that the transfer is too large to produce a two-dimensional superconductor. Interestingly, a similar approach was proposed to e$^-$–dope cuprates in a heterostructure with a manganite[41].

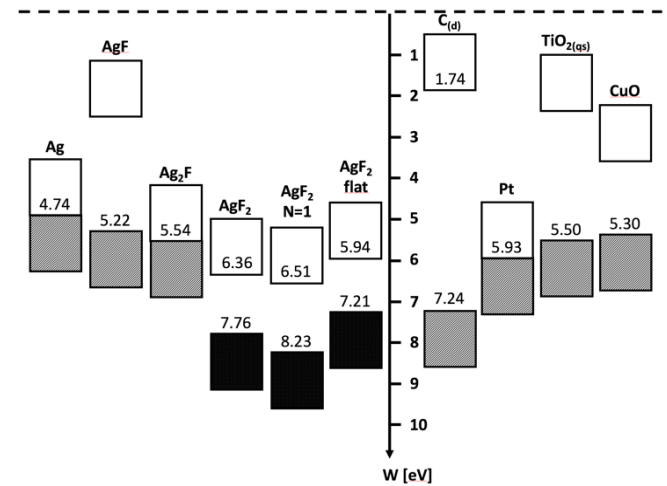

*Figure 4. The calculated positions of the top of the valence band (i.e. work function), as well as the bottom of the conduction band (determined from the calculated band gap) for AgF$_2$ as compared to the values for several other silver systems, as well as Ti(IV) and Cu(II) oxides, Pt metal, and fluorinated surface of diamond (with a single layer of C–F bonds on the surface). The valence band and conduction band were schematically illustrated for all materials by equally-sized rectangles.*

The broken-gap alignment implies that AgF$_2$ is able to inject holes in many wide-band-gap insulators. Such holes would form a two-dimensional hole gas close to the two-dimensional electron gas in AgF$_2$. In this context, a hypothetical interface between a parent cuprate and AgF$_2$ would be particularly intriguing. The workfunction of La$_2$CuO$_4$ is estimated to be 5.02 eV not far from CuO and therefore showing a broken-gap alignment[42]. This may lead to a unique hole-unconventional-superconductor/electron-unconventional-superconductor junction a situation which calls for more research beyond our present scope.

It is of interest to identify metallic substrates that could act as electrodes to study transport properties of AgF$_2$. One candidate could be platinum metal, which is the most difficult to oxidize among elemental metals. Unfortunately, the very high work-function found implies that a contact of $AgF_2$ and a Pt surface would result not only in electron transfer but additionally in F atom transfer from Ag to Pt, and fluorination of the "noble" metal surface. Indeed, it was observed that both Pt and Au form resistive fluoride layers when in contact with Ag(II) fluoride systems[43].

To the best of our knowledge, there are almost no materials with such a high work function as that of $AgF_2$. Recently, it has being proposed[44] that fluorination of borophene, a two-dimensional version of boron grown on a silver substrate, could have a similar high work function. However, the fluorination of such monolayer without affecting the underlying metal would be quite challenging. To the best of our knowledge, the only material for which a similar high work function has been measured is diamond, which has been fluorinated on its (100) surface[45]. Its work function is known to grow from 4.15 eV up to 7.24 eV upon fluorination. This suggests that appropriately doped and conducting material, e.g. boron-doped diamond (BDD) with its parent work function of 4.2 eV[46,47], which has also been prefluorinated on its surface, could constitute a useful conducting substrate, on which a monolayer of $AgF_2$ might be deposited in high-vacuum experiments. The Fermi level difference expected at such junction, would result in only partial injection of electrons into $AgF_2$ layer, thus possibly leading to unconventional superconductivity mediated by antiferromagnetic fluctuations [14–19]. At the same time, the hole doped substrate may display conventional phonon mediated superconductivity of the kind observed in high-pressure synthetized boron doped diamond[48]. This may lead to an electron-unconventional-superconductor / hole-conventional superconductor junction which also deserves more attention.

The fact that the BDD electrode might indeed constitute an appropriate substrate for $AgF_2$, and it would not be chemically deteriorated in contact with this super-potent oxidizer, was confirmed in recent experiments. There, the BDD electrode immersed in anhydrous HF has been successfully applied for demanding macroelectrolysis of $Ag(I)HF_2$ solution resulting in formation for the bulk $Ag(II)F_2$. Only minor deterioration of electrode surface at its most sharp edges was observed in these experiments using SEM[24]. An additional modification of the surface electrode might be desired to deposit a layer of an appropriate metal fluoride[49] on which flat $AgF_2$ layer could be epitaxially grown[40]. Some advantages of the flat layer over the corrugated one have been described above.

In terms of work function and the energy of the bottom of the conduction band, the flat $AgF_2$ layer is the more suited to be hole doped in that it has a reduced work function in respect to the bulk version or a fully relaxed monolayer. It is also the one with the smallest oxidizing power i.e. the least reactive among the three structures. Possible substrates to grow a flat $AgF_2$ layer have been recently proposed[40].

## 4. Conclusions

Our study shows that puckering of the layers is enhanced in a single layer of $AgF_2$ as compared to bulk crystal. This leads to dramatic decrease of the magnetic superexchange constant by an order of magnitude as compared to the bulk system. For bulk $AgF_2$, its calculated work function across the (010) surface reaches 7.76 eV in agreement with the observations that this compound is among the strongest oxidizing agents known. The corresponding value for the relaxed single layer of $AgF_2$ is even larger, 8.23 eV. This implies that very few systems, mostly simple metal fluorides, and fluorinated surface of diamond, could substitute an appropriate substrate for deposition of $AgF_2$ layer(s) without electron transfer at the interphase. Moreover, electron doping constitutes a much more realistic scenario than hole doping. This last possibility may be easier with a flat single $AgF_2$ layer, which is characterized by the smaller work function of 7.21 eV. Appropriate fluoride substrates on which a flat layer of $AgF_2$ could be grown are under examination[40].

The high work functions found enable hole injection in wide-band gap insulators and several oxides. This may lead to intriguing junctions in which a two-dimensional electron gas is in close proximity to a two-dimension hole gas of interest for fundamental studies but also for optoelectronic applications as solar cells[50] and diodes[51].

The current study sets the stage for a realistic design of epitaxially grown $AgF_2$ layers on appropriate substrate materials, as needed to inject charges either from the substrate itself or in a field-effect transistor setup[40], or both.

## Acknowledgements

WG thanx to the Polish National Science Center (NCN) for the Beethoven project (2016/23/G/ST5/04320). K.T. and M.D. acknowledge the European Regional Development Fund, Research and Innovation Operational Program, for project No. ITMS2014+: 313011W085 and Slovak Research and Development Agency, grant No. APVV-18-0168. JL acknowledges financial support from Italian MIUR through Project No. PRIN 2017Z8TS5B, and from Regione Lazio (L. R. 13/08) through project SIMAP. The research was carried out using machines of the Interdisciplinary Centre for Mathematical and Computational Modelling (ICM), University of Warsaw under grant ADVANCE++ (no. GA76-19).

## Author contributions

W.W. carried out calculations with contributions from K.T. and M.D. All authors analyzed the data and contributed to writing of the manuscript. J.L. provided management of various aspects of the project. W.G. conceived and oversaw the whole project.

**ORCID iDs**

W.W. 0000-0001-9765-6222
K.T. 0000-0001-6415-1375
M.D. 0000-0002-4404-0392
J.L. 0000-0001-7426-2570
W.G. 0000-0001-7317-5547

# Electronic Supplementary Information

# Gigantic work function in layered AgF$_2$

W. Wegner,[1,2] K. Tokár[3,4], J. Lorenzana,[5] M. Derzsi,[2,3] W. Grochala[2]*

**Contents.**
**S1. Computational methodology and determination of workfunction.**
**S2. Reference system: N-surface NaF systems slabs from optimized bulk material.**
**S3. Reference system: N-surface AgF systems slabs from optimized bulk material.**
**S4. Reference system: N-surface Ag$_2$F systems slabs from optimized bulk material.**
**S5. Crystallographic Information Files.**
**S6. References.**

## S1. Computational methodology and determination of workfunction:

Projector-augmented-wave (PAW) method was applied[1,2] with rotationally invariant LSDA+U approach introduced by Liechtenstein *et al.*[3]. U and J values for Ag were set to be 5 eV and 1 eV, respectively, as it was previously applied for bulk $AgF_2$ [4,5]. Vienna Ab initio simulation package, VASP 5.4.1, was used for all computations[6–9].

$AgF_2$ bulk: Cut-off energy for plane wave basis set was set to $E_{cut}$ = 520 eV. The blocked Davidson iteration scheme was used. Force, stress tensor, ions, cell shape and cell volume were relaxed. The energy convergence criterion (SCF) was set to 1 x $10^{-7}$ eV. The relaxations were stopped if all forces were smaller than 0.0003 eV/Å. Interpolation formula according to Vosko, Wilk and Nusair was used[10]. Gaussian smearing was used while ionic relaxation with conjugate gradient algorithm were performed. Tetrahedron method with Blöchl corrections were used for no update option. Smearing width was set to 0.04 eV. The k-point mesh of 12x12x12 was used for the magnetic cell identical with unit cell. The optimized lattice parameters of *P*bca cell 5.499 Å, 5.826 Å, and 5.056 Å match well the experimental values of 5.529 Å, 5.813 Å and 5.073 Å, respectively[11], discrepancies not exceeding 0.6%. Figure S2 shows calculated DOS for bulk system.

Unoptimized $AgF_2$ N-layer systems from the bulk: In order to calculate work function of $AgF_2$, N=1, 3, 5, 7, 9, 11 surface systems were constructed from preoptimized bulk $AgF_2$ with their geometry constrained to that found in a crystal. Unit cell were multiplied in the direction perpendicular to the layers to obtain the desired number of layers for further modelling, and to expose the (010) surface sheet towards the vacuum. Such choice of the surface is most natural since the inter-sheet interactions are weak leading to facile cleavage of the layers. *Ca.* 22.5 Å vacuum slab was added to obtained slabs (*cf.* SI for exemplary structure, Figure S1). Single point energies were calculated, followed by the density of states (with DOS evaluation at 3000 grid points) and the total local potential calculations. k-mesh in direction of added vacuum was reduced accordingly to preserve mesh density, with remaining parameters unchanged as compared to the bulk system. Exemplary DOS for N=9 system is showed at Figure S3.

Optimization of $AgF_2$ N=1 layer system: Geometry of the N = 1 $AgF_2$ layer system was optimized. The system was created by first cutting off the middle $AgF_2$ layer from the bulk structure. To prevent the collapse of the vacuum slab during optimization, the void was filled by adding helium atoms in the fractional (0, ½,0) and (½,½,½) sites. The distance between helium atoms was sufficiently large (over twice their van der Waals radius) to prevent their repulsion; in this way, small atoms of totally inert noble gas served as separator, which does not influence the crystal structure. After optimization, helium atoms were removed, and a vacuum slab was increased to ~12.5 Å while preserving the geometry of the $AgF_2$ single layer, and angles between unit cell vectors. So constructed system was used for calculation of the work function.

Optimization of hypothetical $AgF_2$ N=1 flat layer system: Geometry of the N = 1 hypothetical $AgF_2$ flat layer system was optimized. The preliminary structure model was similar to the one described above, albeit it was forced to be tetragonal, with F atoms placed in (½00) and (00½) positions. Similarly, to prevent the collapse of the vacuum slab during optimization, the void was filled by adding helium atoms in the fractional (0,½,0) and (½,½,½) sites. After optimization, helium atoms were removed, and a vacuum slab was increased to ~12.5 Å while preserving the geometry of the $AgF_2$ single flat layer.

AgF reference system:
Bulk AgF was optimized using DFT without spin polarization. The optimized lattice parameter of 4.88 Å matches the experimental value of 4.92 Å[12], discrepancy not exceeding *c.a.* 0.8%. After that N-layered systems for N=1,3, 5, 7, 9, 11 and exposing the (100) surface, were created using preoptimized bulk, and calculations with frozen geometries were performed. AgF slabs were separated with a vacuum one of *ca.* 22.5 Å.

Ag$_2$F reference system:
Bulk Ag$_2$F was optimized in the analogous manner as AgF. The optimized lattice parameter of a=b=2.945 Å and c=5.727 Å matches the experimental value of a=b=2.999 and c= 5.695 Å, discrepancy not exceeding 1.8% and 0.5%, respectively. After that N-layered systems for N=1, 3, 5, 7, 9, 11 and exposing the (001) surface (F terminals) were created in an analogous manner as for AgF, using preoptimized bulk, and calculations with frozen geometries were performed. Ag$_2$F Slabs were separated with a vacuum slab of *ca.* 22.5 Å.

Evaluation of work function:
The minimum work needed to remove one electron from an oriented slab occupied states to a remote point in vacuum is characterised by work function (W) defined by the difference[13]:

$W = E_{vacuum} - E_F$    (Eq.1),

where ($E_{vacuum}$) is electrostatic potential in vacuum near surface as obtained from a maximum of planar averaged local electrostatic potential in supercell vacuum space between slab mirrors (determined using the VASP and P4VASP tool[14], and ($E_F$) is the Fermi level of the slab.

P4VASP and VESTA programs were used for drawings in this work[14,15]

## S2. Reference system: N-surface NaF systems slabs from optimized bulk material:

Bulk NaF was optimized with the same parameters as for bulk AgF$_2$, using DFT without spin polarization. The optimized lattice parameter of 4.628 Å matches the experimental value of 4.619 Å[16], discrepancy not exceeding 0.2%. After that N-layered systems for N=3, 5, 7, 9, 11 and exposing the (100) surface were created in an analogous manner as for AgF$_2$, using preoptimized bulk, and calculations with frozen geometries were performed. Slabs were separated with a vacuum slab of *ca.* 24.5 Å. The same set of calculations were conducted for N-layered NaF systems as previously for unoptimized AgF$_2$ N-layered ones.

In Table 1 values obtained for NaF reference system are summarized. Here we see that with increasing thickness of the slab (N) values of $E_{vacuum}$ and $E_F$ are increasing. Work Function is rather stable and not dependent on slab thickness, and its value can be approximated to 6.7 eV.

*Table 1. Results obtained from DFT calculation for not optimized N-layered systems of NaF build from optimized bulk material: Band Gap (BG), Vacuum energy ($E_{vacuum}$), Fermi Energy ($E_F$) and Work Function (W).*

| N | BG [eV] | $E_{vacuum}$ | $E_F$ | W |
|---|---|---|---|---|
| 3 | 5.648 | 1.090 | -5.580 | 6.671 |
| 5 | 5.719 | 1.560 | -5.121 | 6.681 |
| 7 | 5.745 | 1.917 | -4.766 | 6.683 |
| 9 | 5.761 | 2.198 | -4.483 | 6.681 |
| 11 | 5.762 | 2.424 | -4.262 | 6.686 |
| bulk | 5.553 | --- | -1.993 | --- |

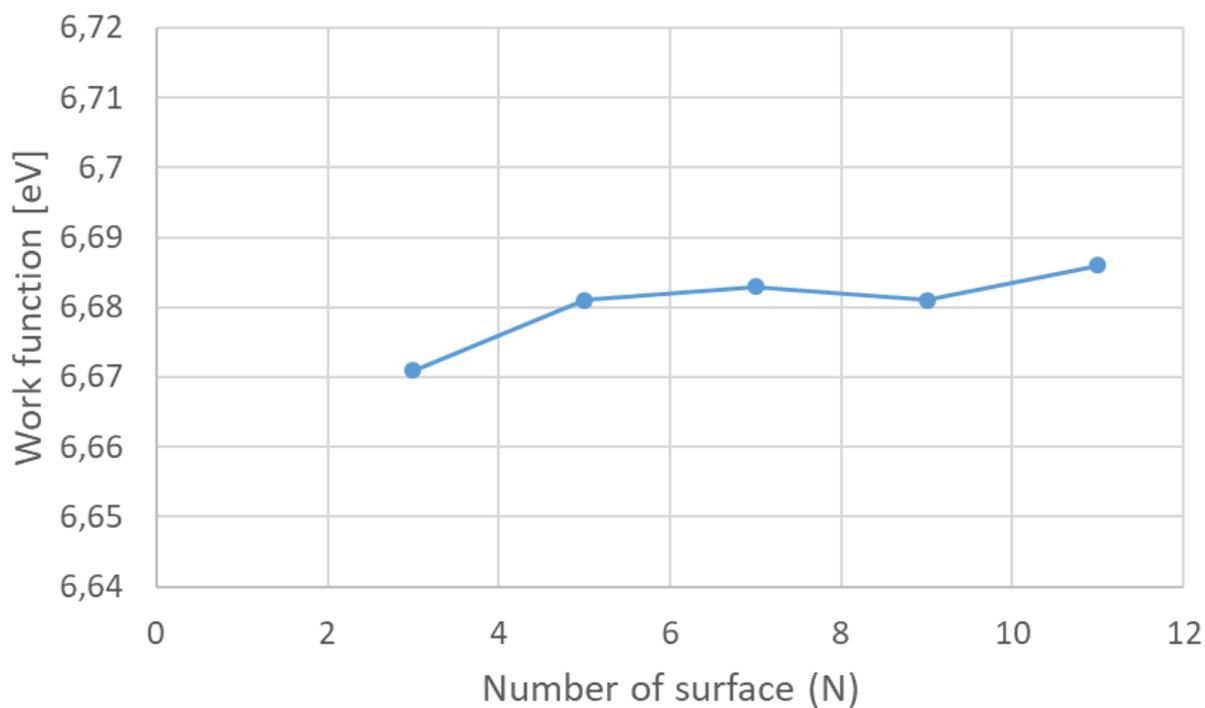

Figure S1. Work function for NaF.

**S3. Reference system: N-surface AgF systems slabs from optimized bulk material:**

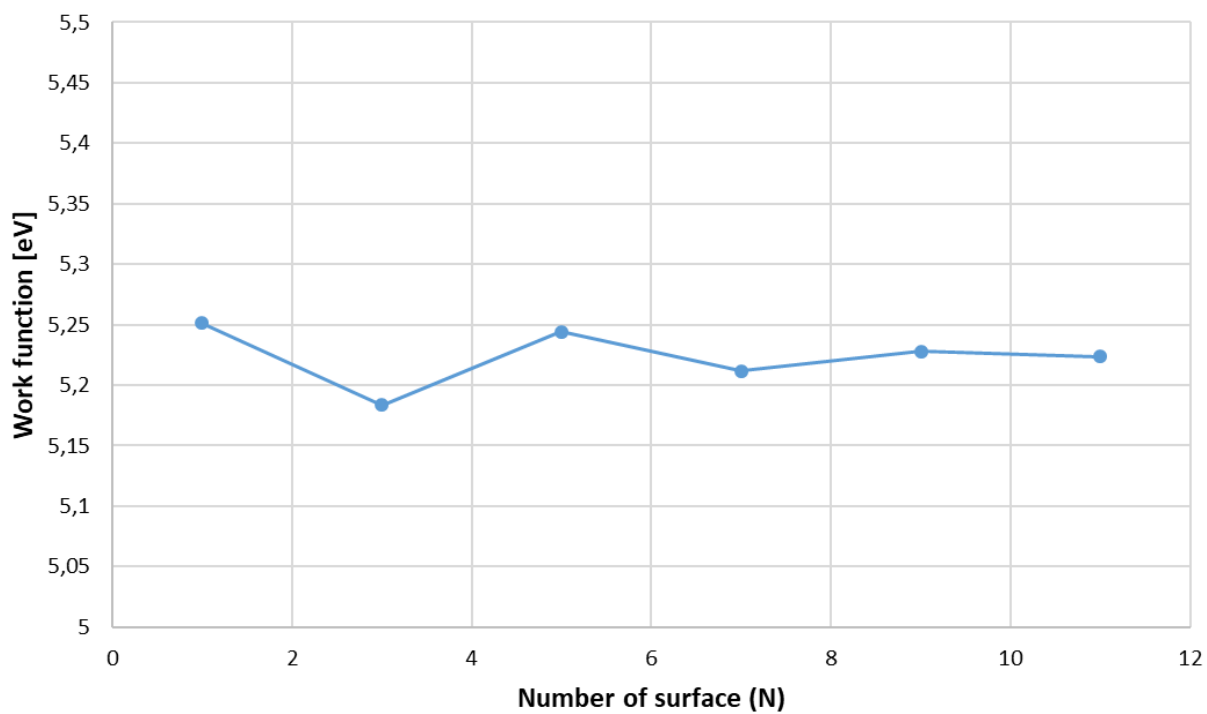

Figure S2. Work function for AgF.

*Table 2. Results obtained from DFT calculation for not optimized N-layered systems of AgF build from optimized bulk material: Vacuum energy ($E_{vacuum}$), Fermi Energy ($E_F$) and Work Function (W).*

| N | $E_{vacuum}$ | $E_F$ | W |
|---|---|---|---|
| 1 | 0.855 | -4.396 | 5.251 |
| 3 | 2.136 | -3.048 | 5.184 |
| 5 | 3.059 | -2.185 | 5.244 |
| 7 | 3.742 | -1.469 | 5.212 |
| 9 | 4.287 | -0.941 | 5.228 |
| 11 | 4.719 | -0.505 | 5.224 |

## S4. Reference system: N-surface Ag$_2$F systems slabs from optimized bulk material:

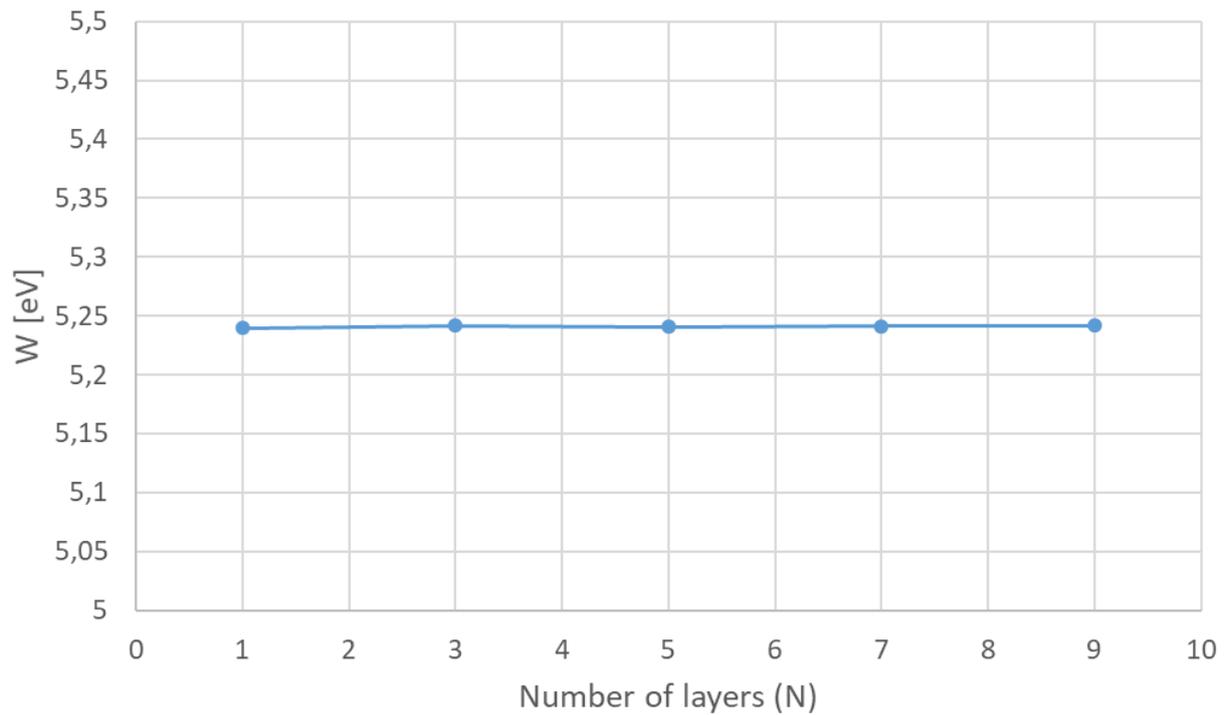

Figure S2. Work function for Ag$_2$F.

*Table 3. Results obtained from DFT calculation for not optimized N-layered systems of NaF build from optimized bulk material: Band Gap (BG), Vacuum energy ($E_{vacuum}$), Fermi Energy ($E_F$) and Work Function (W).*

| N | $E_{vacuum}$ | $E_F$ | W |
|---|---|---|---|
| 1 | 2.093 | -3.147 | 5.240 |
| 3 | 4.639 | -0.602 | 5.242 |
| 5 | 6.121 | 0.880 | 5.241 |
| 7 | 7.087 | 1.846 | 5.241 |
| 9 | 7.768 | 2.526 | 5.242 |

## S5. Crystallographic Information Files.

CIF file for bulk AgF$_2$ optimized system:

#================================================================

# CRYSTAL DATA

#----------------------------------------------------------------

data_AgF2_Bulk

```
_chemical_name_common            ''
_cell_length_a                   5.49911
_cell_length_b                   5.82599
_cell_length_c                   5.05574
_cell_angle_alpha                90
_cell_angle_beta                 90
_cell_angle_gamma                90
_space_group_name_H-M_alt        'P 1'
_space_group_IT_number           1

loop_
_space_group_symop_operation_xyz
 'x, y, z'

loop_
  _atom_site_label
  _atom_site_occupancy
  _atom_site_fract_x
  _atom_site_fract_y
  _atom_site_fract_z
  _atom_site_adp_type
  _atom_site_B_iso_or_equiv
  _atom_site_type_symbol
  Ag1   1.0   0.000000  -0.000000   0.000000   Biso  1.000000  Ag
  Ag2   1.0   0.500000   0.000000   0.500000   Biso  1.000000  Ag
  Ag3   1.0   0.000000   0.500000   0.500000   Biso  1.000000  Ag
  Ag4   1.0   0.500000   0.500000   0.000000   Biso  1.000000  Ag
  F1    1.0   0.305732   0.868176   0.183685   Biso  1.000000  F
  F2    1.0   0.694268   0.131824   0.816315   Biso  1.000000  F
  F3    1.0   0.194268   0.131824   0.683685   Biso  1.000000  F
  F4    1.0   0.805732   0.868176   0.316315   Biso  1.000000  F
  F5    1.0   0.694268   0.368176   0.316315   Biso  1.000000  F
  F6    1.0   0.305732   0.631824   0.683685   Biso  1.000000  F
  F7    1.0   0.805732   0.631824   0.816315   Biso  1.000000  F
  F8    1.0   0.194268   0.368176   0.183685   Biso  1.000000  F
```

CIF file for N=1 layer AgF$_2$ optimized system:

#====================================================================

# CRYSTAL DATA

#----------------------------------------------------------------------

data_AgF2_1N

_chemical_name_common                  ''
_cell_length_a                         5.72758
_cell_length_b                         25.08496
_cell_length_c                         3.72589
_cell_angle_alpha                      90
_cell_angle_beta                       90
_cell_angle_gamma                      74.49773
_space_group_name_H-M_alt              'P 1'
_space_group_IT_number                 1

loop_
_space_group_symop_operation_xyz
 'x, y, z'

loop_
  _atom_site_label
  _atom_site_occupancy
  _atom_site_fract_x
  _atom_site_fract_y
  _atom_site_fract_z
  _atom_site_adp_type
  _atom_site_B_iso_or_equiv
  _atom_site_type_symbol
  Ag1   1.0   0.000000   0.125000   0.500000   Biso   1.000000   Ag
  Ag2   1.0   0.500000   0.125000   0.000000   Biso   1.000000   Ag
  F1    1.0   0.728499   0.086095   0.407047   Biso   1.000000   F
  F2    1.0   0.271501   0.163905   0.592953   Biso   1.000000   F
  F3    1.0   0.771501   0.163905   0.907047   Biso   1.000000   F
  F4    1.0   0.228499   0.086095   0.092953   Biso   1.000000   F

CIF file for N=1 layer AgF$_2$ optimized tetragonal flat layer system:

#========================================================================

# CRYSTAL DATA

#----------------------------------------------------------------------

data_AgF2_1Nflat

_chemical_name_common             ''
_cell_length_a                    8.04142
_cell_length_b                    25.58947
_cell_length_c                    8.04037
_cell_angle_alpha                 90
_cell_angle_beta                  90
_cell_angle_gamma                 90
_space_group_name_H-M_alt         'P 1'
_space_group_IT_number            1

loop_
_space_group_symop_operation_xyz
 'x, y, z'

loop_
  _atom_site_label
  _atom_site_occupancy
  _atom_site_fract_x
  _atom_site_fract_y
  _atom_site_fract_z
  _atom_site_adp_type
  _atom_site_B_iso_or_equiv
  _atom_site_type_symbol
  Ag1   1.0   0.000000   0.109214   0.000000   Biso  1.000000 Ag
  Ag2   1.0   0.000000   0.109214   0.500000   Biso  1.000000 Ag
  Ag3   1.0   0.500000   0.109214   0.000000   Biso  1.000000 Ag
  Ag4   1.0   0.500000   0.109214   0.500000   Biso  1.000000 Ag
  F1    1.0   0.250000   0.109214   0.000000   Biso  1.000000 F
  F2    1.0   0.250000   0.109214   0.500000   Biso  1.000000 F
  F3    1.0   0.750000   0.109214   0.000000   Biso  1.000000 F
  F4    1.0   0.750000   0.109214   0.500000   Biso  1.000000 F
  F5    1.0   0.000000   0.109214   0.250000   Biso  1.000000 F
  F6    1.0   0.000000   0.109214   0.750000   Biso  1.000000 F
  F7    1.0   0.500000   0.109214   0.250000   Biso  1.000000 F
  F8    1.0   0.500000   0.109214   0.750000   Biso  1.000000 F

CIF file for N=11 layer AgF$_2$ unoptimized system:

#========================================================================

# CRYSTAL DATA

#----------------------------------------------------------------------

data_AgF2_N11

_chemical_name_common            ''
_cell_length_a                   5.49911
_cell_length_b                   54.95596
_cell_length_c                   5.05574
_cell_angle_alpha                90
_cell_angle_beta                 90
_cell_angle_gamma                90
_space_group_name_H-M_alt        'P 1'
_space_group_IT_number           1

loop_
_space_group_symop_operation_xyz
 'x, y, z'

loop_
  _atom_site_label
  _atom_site_occupancy
  _atom_site_fract_x
  _atom_site_fract_y
  _atom_site_fract_z
  _atom_site_adp_type
  _atom_site_B_iso_or_equiv
  _atom_site_type_symbol
  Ag1    1.0  0.000000   0.106012   0.000000   Biso  1.000000  Ag
  Ag2    1.0  0.000000   0.212024   0.000000   Biso  1.000000  Ag
  Ag3    1.0  0.000000   0.318036   0.000000   Biso  1.000000  Ag
  Ag4    1.0  0.000000   0.424048   0.000000   Biso  1.000000  Ag
  Ag5    1.0  0.000000   0.530060   0.000000   Biso  1.000000  Ag
  Ag6    1.0  0.500000   0.106012   0.500000   Biso  1.000000  Ag
  Ag7    1.0  0.500000   0.212024   0.500000   Biso  1.000000  Ag
  Ag8    1.0  0.500000   0.318036   0.500000   Biso  1.000000  Ag
  Ag9    1.0  0.500000   0.424048   0.500000   Biso  1.000000  Ag
  Ag10   1.0  0.500000   0.530060   0.500000   Biso  1.000000  Ag
  Ag11   1.0  0.000000   0.053006   0.500000   Biso  1.000000  Ag
  Ag12   1.0  0.000000   0.159018   0.500000   Biso  1.000000  Ag
  Ag13   1.0  0.000000   0.265030   0.500000   Biso  1.000000  Ag
  Ag14   1.0  0.000000   0.371042   0.500000   Biso  1.000000  Ag

| | | | | | | |
|---|---|---|---|---|---|---|
| Ag15 | 1.0 | 0.000000 | 0.477054 | 0.500000 | Biso 1.000000 | Ag |
| Ag16 | 1.0 | 0.000000 | 0.583066 | 0.500000 | Biso 1.000000 | Ag |
| Ag17 | 1.0 | 0.500000 | 0.053006 | 0.000000 | Biso 1.000000 | Ag |
| Ag18 | 1.0 | 0.500000 | 0.159018 | 0.000000 | Biso 1.000000 | Ag |
| Ag19 | 1.0 | 0.500000 | 0.265030 | 0.000000 | Biso 1.000000 | Ag |
| Ag20 | 1.0 | 0.500000 | 0.371042 | 0.000000 | Biso 1.000000 | Ag |
| Ag21 | 1.0 | 0.500000 | 0.477054 | 0.000000 | Biso 1.000000 | Ag |
| Ag22 | 1.0 | 0.500000 | 0.583066 | 0.000000 | Biso 1.000000 | Ag |
| F1 | 1.0 | 0.305732 | 0.092037 | 0.183685 | Biso 1.000000 | F |
| F2 | 1.0 | 0.305732 | 0.198049 | 0.183685 | Biso 1.000000 | F |
| F3 | 1.0 | 0.305732 | 0.304061 | 0.183685 | Biso 1.000000 | F |
| F4 | 1.0 | 0.305732 | 0.410073 | 0.183685 | Biso 1.000000 | F |
| F5 | 1.0 | 0.305732 | 0.516085 | 0.183685 | Biso 1.000000 | F |
| F6 | 1.0 | 0.694268 | 0.119987 | 0.816315 | Biso 1.000000 | F |
| F7 | 1.0 | 0.694268 | 0.225999 | 0.816315 | Biso 1.000000 | F |
| F8 | 1.0 | 0.694268 | 0.332011 | 0.816315 | Biso 1.000000 | F |
| F9 | 1.0 | 0.694268 | 0.438023 | 0.816315 | Biso 1.000000 | F |
| F10 | 1.0 | 0.694268 | 0.544035 | 0.816315 | Biso 1.000000 | F |
| F11 | 1.0 | 0.194268 | 0.119987 | 0.683685 | Biso 1.000000 | F |
| F12 | 1.0 | 0.194268 | 0.225999 | 0.683685 | Biso 1.000000 | F |
| F13 | 1.0 | 0.194268 | 0.332011 | 0.683685 | Biso 1.000000 | F |
| F14 | 1.0 | 0.194268 | 0.438023 | 0.683685 | Biso 1.000000 | F |
| F15 | 1.0 | 0.194268 | 0.544035 | 0.683685 | Biso 1.000000 | F |
| F16 | 1.0 | 0.805732 | 0.092037 | 0.316315 | Biso 1.000000 | F |
| F17 | 1.0 | 0.805732 | 0.198049 | 0.316315 | Biso 1.000000 | F |
| F18 | 1.0 | 0.805732 | 0.304061 | 0.316315 | Biso 1.000000 | F |
| F19 | 1.0 | 0.805732 | 0.410073 | 0.316315 | Biso 1.000000 | F |
| F20 | 1.0 | 0.805732 | 0.516085 | 0.316315 | Biso 1.000000 | F |
| F21 | 1.0 | 0.694268 | 0.039031 | 0.316315 | Biso 1.000000 | F |
| F22 | 1.0 | 0.694268 | 0.145043 | 0.316315 | Biso 1.000000 | F |
| F23 | 1.0 | 0.694268 | 0.251055 | 0.316315 | Biso 1.000000 | F |
| F24 | 1.0 | 0.694268 | 0.357067 | 0.316315 | Biso 1.000000 | F |
| F25 | 1.0 | 0.694268 | 0.463079 | 0.316315 | Biso 1.000000 | F |
| F26 | 1.0 | 0.694268 | 0.569091 | 0.316315 | Biso 1.000000 | F |
| F27 | 1.0 | 0.305732 | 0.066981 | 0.683685 | Biso 1.000000 | F |
| F28 | 1.0 | 0.305732 | 0.172993 | 0.683685 | Biso 1.000000 | F |
| F29 | 1.0 | 0.305732 | 0.279005 | 0.683685 | Biso 1.000000 | F |
| F30 | 1.0 | 0.305732 | 0.385017 | 0.683685 | Biso 1.000000 | F |
| F31 | 1.0 | 0.305732 | 0.491029 | 0.683685 | Biso 1.000000 | F |
| F32 | 1.0 | 0.305732 | 0.597041 | 0.683685 | Biso 1.000000 | F |
| F33 | 1.0 | 0.805732 | 0.066981 | 0.816315 | Biso 1.000000 | F |
| F34 | 1.0 | 0.805732 | 0.172993 | 0.816315 | Biso 1.000000 | F |
| F35 | 1.0 | 0.805732 | 0.279005 | 0.816315 | Biso 1.000000 | F |
| F36 | 1.0 | 0.805732 | 0.385017 | 0.816315 | Biso 1.000000 | F |
| F37 | 1.0 | 0.805732 | 0.491029 | 0.816315 | Biso 1.000000 | F |
| F38 | 1.0 | 0.805732 | 0.597041 | 0.816315 | Biso 1.000000 | F |
| F39 | 1.0 | 0.194268 | 0.039031 | 0.183685 | Biso 1.000000 | F |
| F40 | 1.0 | 0.194268 | 0.145043 | 0.183685 | Biso 1.000000 | F |
| F41 | 1.0 | 0.194268 | 0.251055 | 0.183685 | Biso 1.000000 | F |
| F42 | 1.0 | 0.194268 | 0.357067 | 0.183685 | Biso 1.000000 | F |
| F43 | 1.0 | 0.194268 | 0.463079 | 0.183685 | Biso | 1.000000 F |
| F44 | 1.0 | 0.194268 | 0.569091 | 0.183685 | Biso | 1.000000 F |